\documentclass[11pt]{article}
\usepackage{geometry} 
\usepackage[parfill]{parskip}    
\usepackage{graphicx} 
\usepackage{amssymb}
\usepackage{epstopdf}
\usepackage{alltt} 
\usepackage{rotating} 
\usepackage{rotfloat}
\usepackage{float} 
\usepackage{multirow}
\usepackage{fancyhdr} 
\usepackage{wrapfig}
\usepackage{amsmath}
\usepackage{subfigure}
\usepackage{fullpage}
\usepackage{hyperref}
\usepackage{overcite}
\usepackage{url}
\usepackage{textcomp}
\usepackage{color}
\usepackage[usernames, dvipnames]{xcolor}
\usepackage{gensymb}
\usepackage{wasysym}
\usepackage[utf8]{inputenc}
\usepackage[T1]{fontenc}
\usepackage{braket}

\usepackage{longtable}
\usepackage{booktabs}

\usepackage[symbol]{footmisc}

\usepackage[]{xcolor}

\usepackage[switch]{lineno}

\begin{document}


\begin{center}

\LARGE{\textbf{Low-frequency gravity waves in blue supergiants revealed by high-precision space photometry}}

\end{center}


\vspace{1cm}


Dominic M. Bowman$^{1,}$\footnote[1]{dominic.bowman@kuleuven.be},
Siemen Burssens$^{1}$,
May G. Pedersen$^{1}$,
Cole Johnston$^{1}$,
Conny Aerts$^{1,2}$,
Bram Buysschaert$^{1,3}$,
Mathias Michielsen$^{1}$,
Andrew Tkachenko$^{1}$,
Tamara M. Rogers$^{4,5}$,
Philipp V.~F. Edelmann$^{4}$,
Rathish P. Ratnasingam$^{4}$,
Sergio Sim{\'o}n-D{\' i}az$^{6,7}$,
Norberto Castro$^{8}$,
Ehsan Moravveji$^{1}$,
Benjamin J.~S. Pope$^{9,10}$,
Timothy R. White$^{11}$,
Peter De Cat$^{12}$

\vspace{0.5cm}
	
$^{1}$ Instituut voor Sterrenkunde, KU Leuven, Celestijnenlaan 200D, 3001 Leuven, Belgium \\
$^{2}$ Department of Astrophysics, IMAPP, Radboud University Nijmegen, NL-6500 GL Nijmegen, The Netherlands \\
$^{3}$ LESIA, Observatoire de Paris, PSL Research University, CNRS, Sorbonne Universit{\' e}s, UPMC Univ. Paris 06, Univ. Paris Diderot, Sorbonne Paris Cit{\' e}, 5 place Jules Janssen, F-92195 Meudon, France \\
$^{4}$ School of Mathematics, Statistics and Physics, Newcastle University, Newcastle-upon-Tyne NE1 7RU, UK \\
$^{5}$ Planetary Science Institute, Tucson, AZ 85721, USA \\
$^{6}$ Instituto de Astrof{\' i}sicade Canarias, E-38200 La Laguna, Tenerife, Spain \\ 
$^{7}$ Departamento de Astrof{\' i}sica, Universidad de La Laguna, E-38205 La Laguna, Tenerife, Spain \\
$^{8}$ Leibniz-Institut f{\"u}r Astrophysik Potsdam (AIP), An der Sternwarte 16, 14482, Potsdam, Germany \\
$^{9}$ Center for Cosmology and Particle Physics, Department of Physics, New York University, 726 Broadway, New York, NY 10003, USA \\
$^{10}$ NASA Sagan Fellow \\
$^{11}$ Research School of Astronomy and Astrophysics, Mount Stromlo Observatory, The Australian National University, ACT 2611, Australia \\
$^{12}$ Royal Observatory of Belgium, Ringlaan 3, 1180 Brussel, Belgium




\newpage

\normalsize

{\bf Almost all massive stars explode as supernovae and form a black hole or neutron star. The remnant mass and the impact of the chemical yield on subsequent star formation and galactic evolution strongly depend on the internal physics of the progenitor star, which is currently not well understood. The theoretical uncertainties of stellar interiors accumulate with stellar age, which is particularly pertinent for the blue supergiant phase. Stellar oscillations represent a unique method of probing stellar interiors, yet inference for blue supergiants is hampered by a dearth of observed pulsation modes. Here we report the detection of diverse variability in blue supergiants using the K2 and TESS space missions. The discovery of pulsation modes or an entire spectrum of low-frequency gravity waves in these stars allow us to map the evolution of hot massive stars towards the ends of their lives. Future asteroseismic modelling will provide constraints on ages, core masses, interior mixing, rotation and angular momentum transport. The discovery of variability in blue supergiants is a step towards a data-driven empirical calibration of theoretical evolution models for the most massive stars in the Universe.}


\vspace{0.5cm}

Stars born with masses larger than approximately eight times the mass of the Sun play a significant role in the evolution of galaxies. They are the chemical factories that produce and expel heavy elements through their wind and when they end their lives as supernovae and form a black hole or neutron star \cite{Heger2000a, Maeder2000a, Georgy2013c}{}. However, the chemical yields that enrich the interstellar medium and the remnant mass strongly depend on the progenitor star's interior properties \cite{Nomoto2006}{}. The detectable progenitors of supernovae include blue supergiant stars, which are hot massive stars in a shell-hydrogen or core-helium burning stage of stellar evolution. Stellar evolution models of these post-main sequence stars contain by far the largest uncertainties in stellar astrophysics, as observational constraints on interior mixing, rotation and angular momentum transport are missing. These phenomena are further compounded when coupled with mass loss, binarity and magnetic fields \cite{Heger2000a, Maeder2000a, Georgy2013c}{}. Across astrophysics, from star formation to galactic evolution, it is imperative to calibrate theoretical models of massive stars using observations because they determine the evolution of the cosmos.

A unique methodology for probing stellar interiors is asteroseismology \cite{ASTERO_BOOK}{}, which -- similarly to seismology of earthquakes -- uses oscillations to derive constraints on the structure of stars. The study of stellar interiors of low-mass stars like the Sun has undergone a revolution in recent years because of space telescopes, which have observed thousands of pulsating stars with unprecedented precision and duration. Since oscillations probe the physics of stellar structure, they have successfully been used to distinguish shell-hydrogen and core-helium burning red giant stars \cite{Bedding2011}{}, and to measure interior rotation profiles of thousands of low- and intermediate-mass stars \cite{Beck2012a, Mosser2012c}{}. These discoveries revealed that current angular momentum transport theory is erroneous by at least an order of magnitude \cite{Aerts2019a*}{}.

Asteroseismology of massive stars is missing primarily due to lack of appropriate space telescope observations of these stars. Here we present a large and homogenous sample of hot massive stars observed by the K2 \cite{Howell2014}{} and Transiting Exoplanet Survey Satellite (TESS)\cite{Ricker2015}{} space missions, which includes dozens of blue supergiants. Up to now, few blue supergiants have been shown to pulsate in coherent modes \cite{Saio2006b}{}, as their long-period pulsations are infamously difficult to detect using ground-based telescopes. Furthermore, the opacity driving mechanism responsible for the excitation of coherent pulsation modes in massive stars is strongly dependent on a star’s metallicity, such that low-metallicity massive stars are not expected to pulsate in such modes \cite{Salmon_S_2012}{}. In addition to coherent oscillations, it remains unknown whether a ubiquitous cause of stochastic low-frequency variability exists in massive stars, with the interaction of photospheric and wind variability seeming to play an important role \cite{Aerts2009b, Aerts2017a, Simon-Diaz2018a}{}. Therefore, detailed scientific inference had to await high-quality, uninterrupted and long-duration space photometry, which is vital for unravelling and interpreting the complex variability of hot massive stars \cite{Aerts2017a}{}.

To test the incidence of coherent pulsation modes and stochastic low-frequency variability in hot massive stars, we assembled a sample of 114 ecliptic stars and 53 Large Magellanic Cloud (LMC) stars of spectral type O or B with available K2 and TESS space photometry, respectively. We refer to Supplementary Tables~1 and 2 for the names and spectral types of all stars. Our K2 sample comprises stars from various ecliptic campaigns; their light curves have a cadence of 30~min and are up to 90~d in length. We created optimum pixel masks \cite{Buysschaert2015}{}, and performed robust detrending of instrumental signals using Gaussian processes with the \texttt{k2sc} software package \cite{Aigrain2016a}{}. The TESS light curves of the LMC sample have a 2-min cadence and are from sectors 1--3 and are similarly up to 90~d in length. These data reveal variability with time scales between a few hours and several weeks, and amplitudes of order 10~$\mu$mag owing to the unprecedented precision of space photometry. We identified significant frequencies in the amplitude spectra of the K2 stars indicative of coherent pulsation modes using iterative pre-whitening \cite{ASTERO_BOOK}{}, in which the sinusoidal model
\begin{equation}
\Delta m = A \cos(2\pi\nu(t - t_0)+\phi) ~ ,
\label{equation: sinusoid}
\end{equation}
is fitted and subtracted from the light curve of time $t$ normalized to a zero-point $t_0$, using a frequency $\nu$, an amplitude $A$ and a phase $\phi$. The details of our analysis are given in the Methods.

\subsection*{Results}

The independent K2 and TESS samples allow a unique insight into the diversity of pulsations in massive stars for different stellar metallicities. The K2 ecliptic sample revealed 37 of the 114 stars to have coherent pressure and/or gravity pulsation modes, which includes several previously unknown Beta Cephei ($\beta$~Cep) stars. The TESS sample did not reveal any stars with high-amplitude multi-periodic coherent pulsation modes, thus no iterative pre-whitening of high-amplitude signals was necessary to probe the morphology of stochastic low-frequency variability in these stars. The TESS data support the dearth of opacity-driven coherent pulsation modes predicted for massive stars within the low-metallicity environment of the LMC \cite{Salmon_S_2012}{}.

To characterize stochastic low-frequency variability in a star, we subsequently used a Bayesian Markov chain Monte Carlo framework using the \texttt{Python} code \texttt{emcee} \cite{Foreman-Mackey2013}{} to fit an amplitude spectrum with the function
\begin{equation}
\alpha \left( \nu \right) = \frac{ \alpha_{0} } { 1 + \left( \frac{\nu}{\nu_{\rm char}} \right)^{\gamma}} + C_{\rm w} ~ ,	
\label{equation: red noise}
\end{equation}
\noindent where $\alpha_{0}$ represents the amplitude at a frequency of zero, $\gamma$ is the logarithmic amplitude gradient, $\nu_{\rm char}$ is the characteristic frequency, which is the inverse of the characteristic timescale, $\tau$, of stochastic variability present in the light curve such that $\nu_{\rm char} = (2\pi\tau)^{-1}$, and $C_{\rm w}$ is a frequency-independent (white) noise term \cite{Blomme2011b, Bowman2019a}{}. The validity of the measured fit using equation~(\ref{equation: red noise}) is determined using the Bayesian information criterion in comparison to a linear white noise model. Within the K2 sample, 3 of the 114 stars are found to have only white noise in their amplitude spectra after removal of coherent pulsation modes, if any. Therefore, 111 of the K2 stars and all 53 of the TESS LMC stars reveal significant low-frequency variability in their amplitude spectra, which is indicative of stochastic photospheric variability in the form of an entire spectrum of internal gravity waves (IGWs) and/or rotational modulation by a clumpy aspherical wind \cite{Aerts2017a, Simon-Diaz2018a, Bowman2019a}{}. The detected spectra of IGWs also contain standing waves, which correspond to the gravity modes of the star \cite{Aerts2019a*}{}. We refer to Supplementary Tables~3 and 4 for the individual fit values using equation~(\ref{equation: red noise}) to each star.

\begin{figure}[H]
\centering
\includegraphics[]{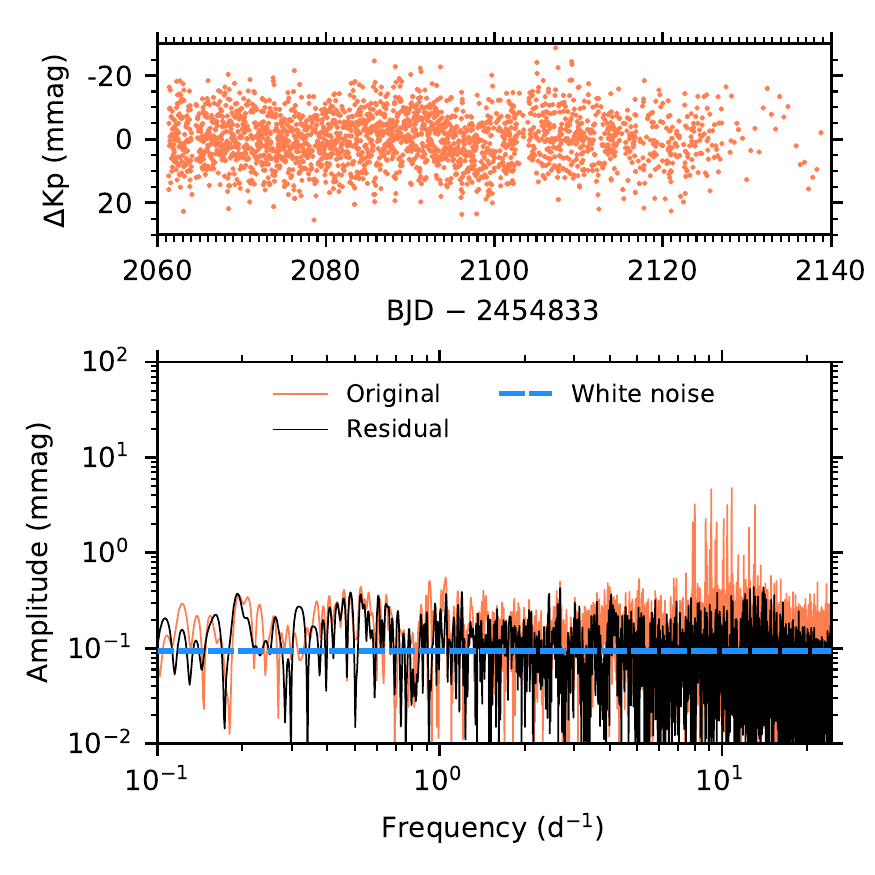}
\caption{{\bf K2 data of the $\beta$~Cep star EPIC~202929357.} The light curve is shown in the top panel, and the corresponding original and residual amplitude spectra are shown in orange and black, respectively, in the bottom panel. The light curve is given as the change in the Kepler passband magnitude ($\Delta$Kp) as a function of Barycentric Julian date. The coherent pressure-mode oscillation frequencies are between 6 and 12~d$^{-1}$ and the white noise background is indicated by the dashed-blue line.}
\label{figure: example white}
\end{figure}

\begin{figure}[H]
\centering
\includegraphics[]{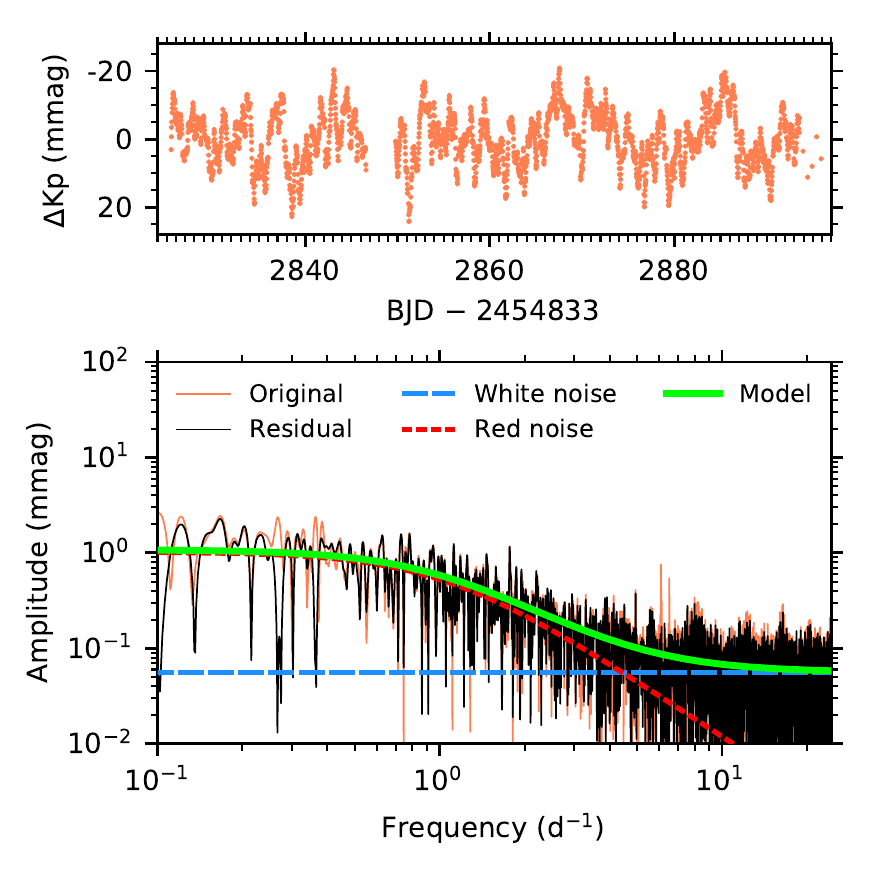}
\caption{{\bf K2 data of the blue supergiant star EPIC~240255386.} The light curve is shown in the top panel, and the corresponding original and residual amplitude spectra are shown in orange and black, respectively, in the bottom panel. The light curve is given as the change in the Kepler passband magnitude ($\Delta$Kp) as a function of Barycentric Julian date. The stochastic low-frequency variability, which is characterized using equation~(\ref{equation: red noise}) and is indicative of photospheric variability caused by gravity waves, is shown as the solid green line, and its red and white components are shown as red- and blue-dashed lines, respectively.}
\label{figure: example red}
\end{figure}

To demonstrate the diversity in variability within our samples, we show the light curves and amplitude spectra of two representative K2 stars in Figs~\ref{figure: example white} and \ref{figure: example red}. The original and residual amplitude spectra after coherent modes have been removed via pre-whitening (cf. equation~\ref{equation: sinusoid}) in these stars are shown in orange and black, respectively. In Fig.~\ref{figure: example white} we present the reduced K2 light curve and amplitude spectrum of the $\beta$~Cep star EPIC~202929357, which has multi-periodic coherent pressure-mode oscillations. This pulsating star is one of the three stars without low-frequency variability in its amplitude spectrum and demonstrates the quality of K2 data. The second example in Fig.~\ref{figure: example red} is a spectroscopically confirmed blue supergiant, EPIC~240255386 with a spectral type of B0\,Ia, which like other blue supergiants in our sample, has stochastic low-frequency photometric variability that was unknown so far. For all the 111 K2 stars with significant low-frequency variability, we characterize the residual amplitude spectrum using equation~(\ref{equation: red noise}) with the best-fit model and its white and red components shown as solid green, and blue- and red-dashed lines, respectively. As none of the 53 TESS LMC stars underwent iterative pre-whitening of coherent pulsation modes, we characterize the morphology of stochastic low-frequency background in these stars using the amplitude spectrum of their original light curves using equation~(\ref{equation: red noise}). We provide the light curves and fitted amplitude spectra for all K2 stars as Supplementary Figs~4--117 and as Supplementary Figs~118--170 for all TESS stars.

\begin{figure}[H]
\centering
\includegraphics[]{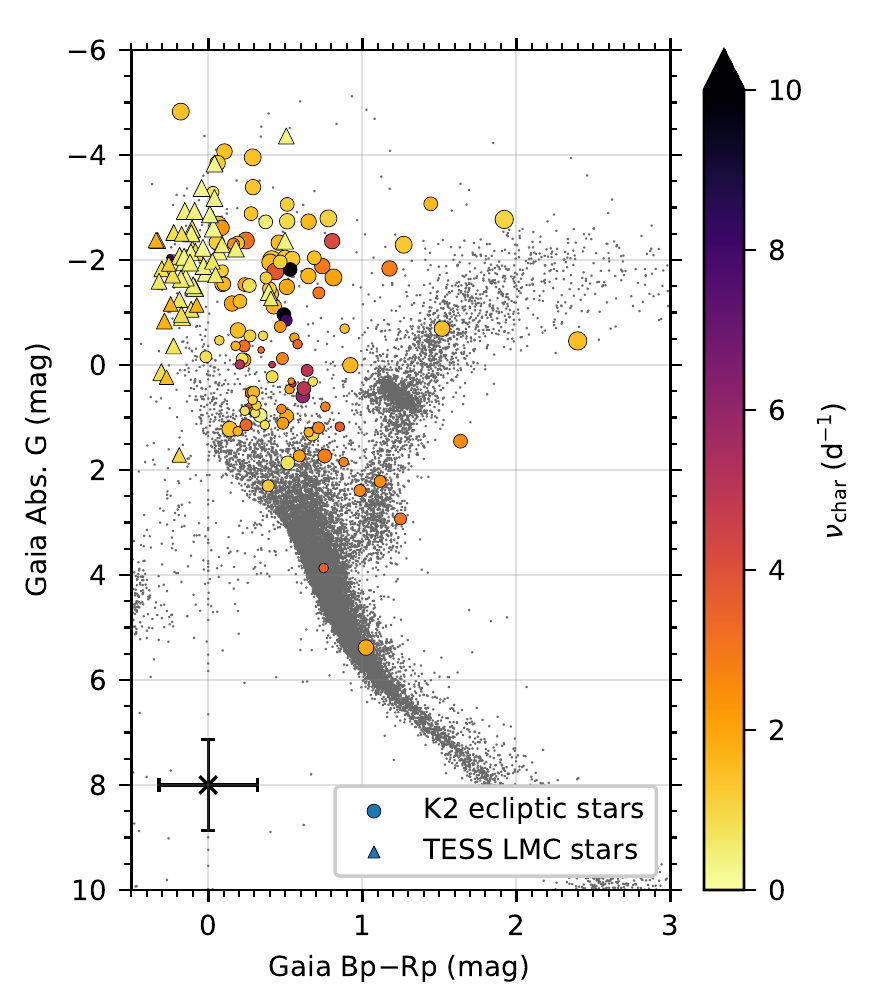}
\caption{{\bf Gaia colour-magnitude diagram of OB stars observed by the K2 and TESS space missions.} The filled circles represent the 111 of 114 and OB stars observed by K2 and the triangles represent the 53 LMC OB stars observed by TESS, which were all concluded to have significant low-frequency variability. All symbols have been colour-coded according to a star's characteristic frequency, $\nu_{\rm char}$, and a symbol size proportional to the amplitude of the stochastic low-frequency variability, $\alpha_0$ (cf. equation~\ref{equation: red noise}). The Gaia colour (Bp--Rp) is the difference between the stellar magnitude measured in the Gaia blue and red passbands. A representative 1-$\sigma$ uncertainty for the location of the OB-star sample is shown in the bottom left corner of the plot. The grey symbols show the distribution of TESS targets in sectors 1--3 that were also observed by Gaia for comparison.}
\label{figure: CMD}
\end{figure}

We summarize the discovery of the pulsational variability in Fig.~\ref{figure: CMD}{}, in which the distribution of our stars in a colour-magnitude diagram using the second data release (DR2) of the Gaia mission \cite{Gaia2018a}{} is shown. We used Gaia distance estimates \cite{Bailer-Jones2018c}{} to calculate absolute Gaia G-band magnitudes, with the location of the K2 and TESS samples in Fig.~\ref{figure: CMD} given as the filled circles and triangles, respectively, colour-coded by the characteristic frequency, $\nu_{\rm char}$, and with a symbol size proportional to the amplitude of the measured low-frequency variability, $\alpha_0$ (cf. equation~\ref{equation: red noise}). We refer to Supplementary Tables~1 and 2 for the Gaia photometry of all stars. For comparison, all the targets observed by the TESS mission in sectors 1--3 have also been plotted in Fig.~\ref{figure: CMD} as grey dots. We also investigated the effect of extinction and reddening for our samples \cite{Green_G_2018a, McCall2004b}{}, with a representative 1-$\sigma$ error bar given in Fig.~\ref{figure: CMD} indicating the typical uncertainty in each star's location in the colour-magnitude diagram given uncertainties in its Gaia distance \cite{Green_G_2018a}{}, reddening and extinction \cite{McCall2004b}{}. The location of the stars in Fig.~\ref{figure: CMD} supports a spectral type of O or B for the vast majority of stars. 

We measure different ranges in $\alpha_0$ and $\nu_{\rm char}$ values within the K2 and TESS samples (we refer to the Supplementary Tables 3 and 4, and Supplementary Figs~1--3), but similar values of the logarithmic amplitude gradient $\gamma$. These findings point to one common physical origin for the excitation of the stochastic low-frequency variability found in O and B stars, with $\alpha_0$ and $\nu_{\rm char}$ dependent on the stellar parameters such as luminosity, mass, and radius. Our K2 sample represents the first homogenous ensemble of high-quality space photometry of hot massive stars in different stages of stellar evolution for which a significant fraction pulsate in coherent pulsation modes. Furthermore, the presence of stochastic low-frequency variability in both our K2 and TESS samples demonstrates this phenomenon is nearly ubiquitous in space photometry of massive stars.

\begin{figure}[H]
\centering
\includegraphics[]{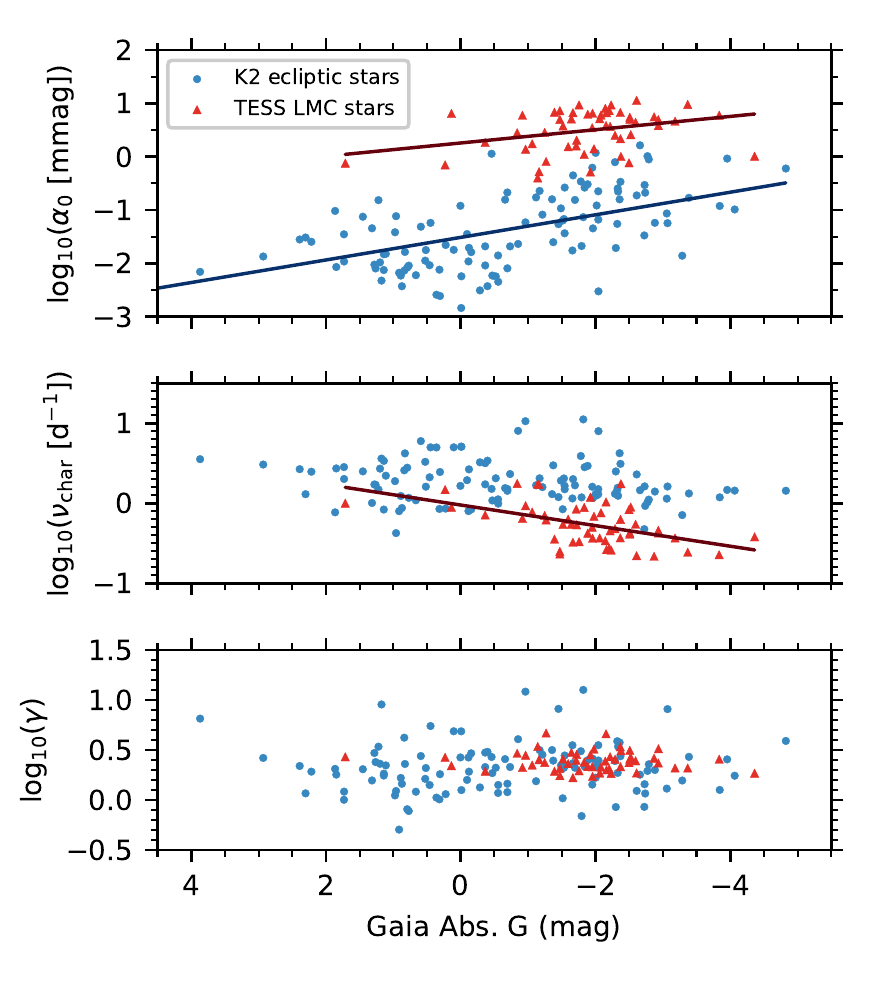}
\caption{{\bf Relationship between intrinsic stellar brightness and IGW morphology.} The pairwise correlation between the Gaia absolute G-band magnitudes and the fit parameters $\alpha_0$, $\nu_{\rm char}$ and $\gamma$ for the K2 and TESS samples are shown as blue circles and red triangles, respectively. Significant correlations ($p < 0.02$) determined from a two-tailed linear regression are shown as solid lines for each sample.}
\label{figure: linear regression}
\end{figure}

Most importantly, we note the trend between the brighter stars (that is more negative absolute magnitudes) and larger amplitudes in the observed stochastic low-frequency variability, $\alpha_0$, which is most apparent within the TESS LMC sample as shown in Fig.~\ref{figure: linear regression}. Linear regression using a two-tailed hypothesis test reveals a significant correlation ($p < 0.02$) between the intrinsic brightness of a star and the amplitude of its low-frequency variability in both samples. Furthermore, a significant correlation ($p < 0.0001$) is also discovered between the intrinsic brightness of a star and the characteristic frequency, $\nu_{\rm char}$ in the 53 TESS LMC stars. This supports the interpretation that the stochastic low-frequency variability observed in OB stars is evidence of IGWs, since the amplitudes of individual waves caused by core convection are predicted to scale with the local core luminosity, and hence mass of the convective core \cite{Rogers2013b, Rogers2015, Edelmann2019a}{}. These results also demonstrate that the low metallicity environment of the LMC stars does not strongly impact the observed morphology of the stochastic low-frequency variability. This is indicative of gravity waves triggered by core convection, opposed to damped pulsation modes driven by an opacity mechanism or waves triggered by thin convective layers due to an opacity peak in the stellar envelope \cite{Cantiello2009a}{}. The individual distributions of the fit parameters, $\alpha_0$, $\nu_{\rm char}$ and $\gamma$ are provided as Supplementary Figs~1, 2 and 3, respectively.

\subsection*{Discussion}

Stochastic variability in the form of low-frequency IGWs excited at the interface of convective and radiative regions is predicted for massive stars \cite{Cantiello2009a, Rogers2013b}{}, but the detection of such waves remained illusive before the advent of space photometry \cite{Aerts2017a, Bowman2019a}{}. These IGWs are efficient at transporting angular momentum \cite{Rogers2013b, Rogers2015}{} and may explain the observed near-rigid rotation profiles in intermediate-mass stars \cite{Aerts2019a*}{}. However there were few firm detections of IGWs in massive stars before this study \cite{Aerts2017a, Simon-Diaz2018a, Bowman2019a}{}. Recent two-dimensional (2D) and 3D spherical numerical simulations using a realistic stellar structure model predict the excitation of IGWs by turbulent core convection, and an amplitude spectrum for IGWs with a value between 0.8 and 3 for the frequency exponent gamma \cite{Rogers2013b, Rogers2015, Edelmann2019a}{}. On the other hand, 3D simulations in a Cartesian box in idealized conditions predict a steeper IGW amplitude spectrum with 3.25 as the frequency exponent \cite{Couston2018b}{}. Our analysis leads to $\gamma \leq 3.5$ for the vast majority of O and B stars. Therefore, the range in measured $\alpha_0$ and $\nu_{\rm char}$ values along with the similar $\gamma$ values from the observations of O and B stars are consistent with the spherical simulations of IGWs caused by core convection. No other physical causes of variability are currently known to result in amplitude spectra with the observed characteristics for the wide range of metal-rich and metal-poor stars presented here.

As a next step, spectroscopy of these stars will allow the determination of fundamental stellar parameters, which are necessary to perform asteroseismic modelling and to constrain interior properties, including the measurement of a star’s age, core mass and interior rotation profile. Future identification of coherent modes, damped modes and/or the standing waves occurring within the observed spectrum of IGWs in terms of their radial order, angular degree and azimuthal order, as it has been achieved from frequency patterns in space photometry \cite{Aerts2019a*}{} or from high-resolution spectroscopy \cite{ASTERO_BOOK}{}, will allow us to distinguish between shell-hydrogen and core-helium burning massive stars. Indeed, gravity modes are sensitive to the core properties, such as the core mass and whether the star has a convective or radiative core. We demonstrate this future potential using an illustrative example of four different massive star models in Fig.~\ref{figure: modelling}, in which zonal dipole mode frequencies for radial orders $n \in [-50, -1]$ and $n \in [1, 10]$ are shown as solid and dashed vertical lines, respectively. The modes of each stellar model differ in frequency range because of the difference in core properties. The details of our model calculations are given in the Methods. As p-modes are mostly sensitive to the outer layers and the global average properties of a star, consecutive radial order p-modes provide a constraint on the mean stellar density. Conversely, g-modes are mostly sensitive to the near-core region and in particular the core mass, and consecutive radial order g-modes of the same angular degree, which is known as a period spacing pattern, provide constraints on the near-core rotation. Asteroseismic modelling provides the most likely estimate of the model parameters by fitting the observed identified pulsation-mode frequencies of a star to theoretical stellar evolution models and their associated predicted eigenfrequencies, while taking advantage of spectroscopic constraints of the effective temperature and surface gravity and/or the luminosity of a star to vastly reduce the parameter space to be considered for models in the Hertzsprung--Russell diagram.

The mode frequencies shown in Fig.~\ref{figure: modelling} were computed in the adiabatic approximation, which is a valid approach for asteroseismic modelling of OB-type pulsators\cite{Aerts2018b}{}. The exploitation of detected oscillation frequencies can be done irrespective of the excitation of the modes and waves. At present, at least four excitation mechanisms are known to trigger waves in OB stars: coherent pressure or gravity modes excited by an envelope heat-mechanism\cite{Szewczuk2017a}{}, stochastic wave generation by the interface layer between the convective core and the radiative envelope\cite{Rogers2013b}{}, stochastically triggered waves by thin sub-surface convection zones in the otherwise radiative envelope due to local opacity enhancements associated with iron and helium ionisation\cite{Cantiello2009a}{}, or tidal excitation\cite{Fuller2017c}{}. All these mechanisms may lead to the detection of the periods of standing gravity waves in the regime shown in Fig~\ref{figure: modelling}, where the bottom panel reveals the period spacing patterns for the four different stellar models along with a typical uncertainty for data from the TESS mission after 1-year of monitoring (from its Continuous Viewing Zone, CVZ\cite{Ricker2015}{}). It can be seen that the future TESS CVZ data will provide us with the opportunity to unravel the evolutionary status of a blue supergiant once the degrees of the standing waves have been identified from combined space photometry and high-precision ground-based spectroscopy following a similar methodology that has been successfully applied to pulsating F- and B-type dwarf stars \cite{VanReeth2016a, Papics2017a}.

Hence, it is the combination of photometry, spectroscopy, astrometry and future forward seismic modelling that will allow the determination of which stars are undergoing blue loops in their evolution. Our sample of OB stars observed by K2 and TESS, which includes many blue supergiants, offers the unique opportunity for asteroseismic interpretation of post-main sequence stars that may be undergoing blue loops in the Hertzsprung--Russell diagram, because their interior structures and core masses are different than those of the stars that do not undergo loops and gravity modes directly probe this \cite{Aerts2019a*}{}. The identification of standing waves within the measured low-frequency morphologies \cite{Edelmann2019a}{} across a range of masses and evolutionary stages are also essential to calibrate and constrain numerical simulations of convectively driven waves in terms of wave excitation, propagation and dissipation. Furthermore, these observational results provide a strong practical guideline for the inclusion of angular momentum transport by IGWs in the next generation of stellar structure and evolution models.


\newcommand{\apj}{\it Astrophys. J.}
\newcommand{\aap}{\it Astron. Astrophys.}
\newcommand{\mnras}{\it Mon. Not. R. Astron. Soc.}
\newcommand{\nat}{\it Nature}
\newcommand{\araa}{\it Annu. Rev. Astron. Astrophys.}
\newcommand{\pasp}{\it Pub. Astron. Soc. Pac.}
\newcommand{\apjl}{\it Astrophys. J.. Letters.}
\newcommand{\aj}{\it Astron. J.}
\newcommand{\apjs}{\it Astrophys. J. Supp. Ser.}




\subsection*{Correspondence}
Correspondence and requests for materials should be addressed to the corresponding author Dominic M. Bowman (dominic.bowman@kuleuven.be).

\subsection*{Acknowledgements}
The K2 and TESS data presented in this paper were obtained from the Mikulski Archive for Space Telescopes (MAST). Funding for the K2 mission is provided by NASA’s Science Mission Directorate. Funding for the TESS mission is provided by the NASA Explorer Program. STScI is operated by the Association of Universities for Research in Astronomy, under NASA contract NAS5-26555. Support for MAST for non-HST data is provided
by the NASA Office of Space Science via grant NNX09AF08G and by other grants and contracts. The Gaia data in this paper come from the European Space Agency mission Gaia, processed by the Gaia Data Processing and Analysis Consortium (DPAC). Funding for the DPAC has been provided by national institutions, in particular the institutions participating in the Gaia Multilateral Agreement. This research has made use of the SIMBAD database, operated at CDS, Strasbourg, France; the SAO/NASA Astrophysics Data System; and the VizieR catalogue access tool, CDS, Strasbourg, France. The research leading to these results has received funding from the European Research Council (ERC) under the European Union’s Horizon 2020 research and innovation programme (grant agreement number 670519: MAMSIE). T.M.R., P.V.F.E. and R.P.R. received support from STFC grant ST/L005549/1 and NASA grant NNX17AB92G. S.S.-D. acknowledges financial support from the Spanish Ministry of Economy and Competitiveness (MINECO) through grants AYA2015-68012-C2-1 and Severo Ochoa SEV-2015-0548, and grant ProID2017010115 from the Gobierno de Canarias. B.J.S.P. is a NASA Sagan Fellow. This work was performed in part under contract with the Jet Propulsion Laboratory funded by NASA through the Sagan Fellowship Program executed by the NASA Exoplanet Science Institute. T.R.W. acknowledges the support of the Australian Research Council (grant DP150100250).


\subsection*{Author Contributions}
Definition of the science case and submission of the K2/TESS guest observer/investigator proposals: C.A., E.M., P.D.C., B.B., M.G.P., A.T., D.M.B., C.J., T.M.R., P.V.F.E. and R.P.R. Processing and data reduction of K2/TESS photometry: D.M.B., S.B., B.B., B.J.S.P. and T.R.W. Frequency analysis, fitting and interpretation of the results: D.M.B., C.J., M.M., C.A. and A.T. Reduction, analysis and interpretation of Gaia photometry: D.M.B.
and M.G.P. Analysis, interpretation and comparison of results to hydrodynamical simulations: T.M.R., P.V.F.E. and R.P.R. Co-ordination of future follow-up spectroscopy: D.M.B., S.S.-D., N.C. and S.B. All authors discussed and commented on the manuscript.


\subsection*{Competing interests }
The authors declare no competing interests.


\newpage 


\begin{figure}[H]
\centering
\includegraphics[width=0.92\textwidth]{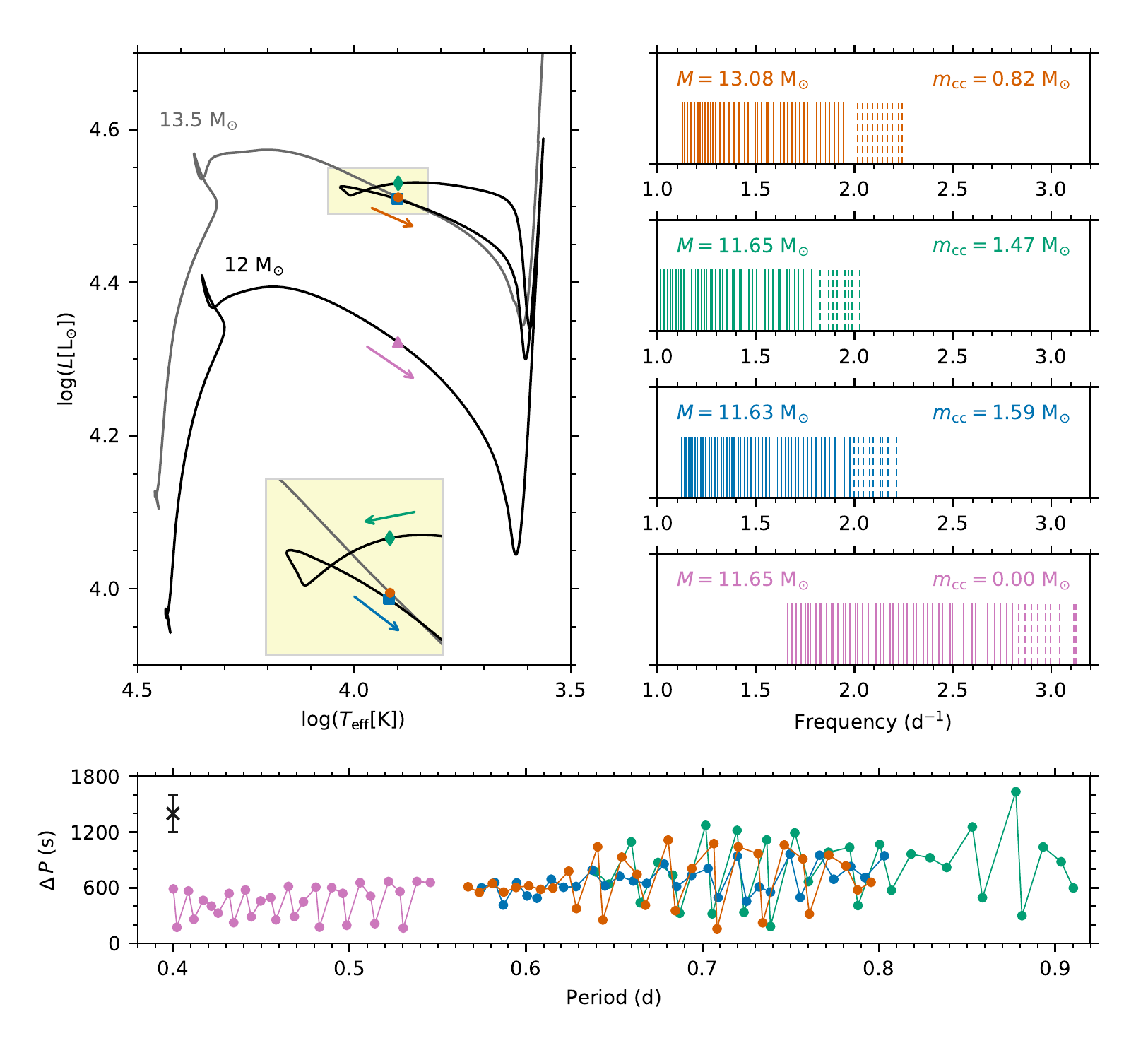}
\caption{{\bf Asteroseismic potential of hot massive post-main sequence stars.} \small In the right-hand panels, the adiabatic dipole zonal mode frequencies for radial orders $n \in [-50, -1]$ and $n \in [1, 10]$ are shown as solid and dashed vertical lines, respectively, for four representative models at different evolutionary stages of stars of initial masses (M) of 12 and 13.5~M$_{\odot}$, whose location in the Hertzsprung--Russell diagram is shown in the left panel ($T_{\rm eff}$ is the stellar effective temperature; $L$ is the stellar luminosity). Identification of standing wave frequencies in terms of their radial order, angular degree and azimuthal order allow stars that may be undergoing blue loops to be distinguished because their interior structures and convective core masses ($m_{\rm cc}$) are significantly different and the oscillation frequencies directly probe this. The arrows show the direction of each star's travel along its respective evolutionary path. The bottom panel shows period spacing patterns of the gravity modes with radial orders in the range $n \in [-40, -10]$. These patterns are defined as the differences between the periods of gravity modes with the same angular degree and consecutive radial order ($\Delta\,P$) as a function of mode periods. The patterns are shown for the same four models, with a typical 1-$\sigma$ uncertainty for observed frequencies also provided in the top left part of the panel (note the uncertainty in the x-axis is smaller than the symbol size).}
\label{figure: modelling}
\end{figure}


\normalsize

\section*{Methods}
\label{section: methods}


\subsection*{Iterative pre-whitening of the K2 and TESS samples}
\label{subsection: K2 sample}

For stars in the K2 sample, we performed iterative pre-whitening of periodic variability until the amplitude signal-to-noise ratio (S/N) dropped below four. At each iteration, the frequency of highest amplitude in the spectrum was used as input to a linear and subsequent nonlinear least-squares fit to the light curve, with the optimized sinusoidal model using equation~(\ref{equation: sinusoid}) subtracted from the light curve to calculate a residual amplitude spectrum. This process was then repeated iteratively until S/N < 4 was reached for all frequencies in the residual light curve. The amplitude S/N value was calculated as a ratio of the amplitude of the extracted frequency to the average of the residuals in the amplitude spectrum within a symmetric frequency window of 1~d$^{-1}$ around the extracted frequency. This conservative approach to pre-whitening (as opposed to calculating the noise using a window at high frequency) ensured that we were not over-fitting or over-extracting the signal via pre-whitening. On average, between 10 and 15 frequencies were extracted for each star, which are not large numbers considering the precision and duration of K2 space mission photometry. We used this approach to characterize the morphology of the stochastic low-frequency background in the residual amplitude spectra of the K2 sample of OB stars, as many of these ecliptic stars exhibit periodic variability in the form of coherent pulsation modes and/or rotation modulation.

In contrast, the LMC sample of OB stars observed by TESS did not undergo pre-whitening as the vast majority of these stars did not reveal high-amplitude multi-periodic pulsation modes; their light curves and amplitude spectra are dominated by stochastic low-frequency variability. A similar distribution in the morphology of the stochastic low-frequency variability is found for both K2 and TESS samples (see Supplementary Figs~1, 2 and 3). Yet the TESS LMC stars exhibit higher amplitudes and these stars are intrinsically brighter on average. Therefore, based on the similar morphologies obtained in the K2 and TESS samples, we concluded that our K2 results are not influenced by the iterative pre-whitening of their coherent pulsation modes.

	
\subsection*{Low-frequency morphology distributions}
\label{subsection: distribution}

We provide the histograms of the fit parameters $\alpha_0$, $\nu_{\rm char}$ and $\gamma$ used to characterize the morphology of the stochastic low-frequency variability in the residual amplitude spectra of 111 K2 ecliptic stars and the original amplitude spectra of 53 TESS LMC stars in Supplementary Figs~1--3, respectively. In these figures, only half of the width of the true bin size is shown, which includes a horizontal offset between the K2 and TESS samples, for clarity. Although there is a range in measured $\alpha_0$, $\nu_{\rm char}$ and $\gamma$ values in our sample of hot massive stars, we found that a common morphology is present such that the amplitude spectra of all stars are well characterized by equation~(\ref{equation: red noise}). This demonstrates that an astrophysical source of stochastic low-frequency variability is present in their light curves, which is indicative of gravity waves. 

We note that the $\log_{10}(\alpha_0)$ and $\log_{10}(C_{\rm W})$ values for both K2 and TESS samples are also correlated, such that for intrinsically brighter stars, $\alpha_0$ and $C_{\rm W}$ are both larger. However, the white noise component in the LMC stars, which are intrinsically bright but observed by TESS to be faint is, on average, larger than that of the K2 sample of stars. The white noise component, which dominates at high frequency in an amplitude spectrum, is typically governed by the photon noise and the data quality, such that stars with a fainter apparent brightness usually have a larger white noise contribution. Given the instrumental differences in the light curves of each sample and large differences in the apparent magnitudes of LMC and K2 stars, we kept the two samples separate in this work.

	
\subsection*{Stellar evolution models}
\label{subsection: models}

To create the illustrative example of the asteroseismic potential of our blue supergiant sample shown in Fig.~\ref{figure: modelling}, we computed non-rotating evolutionary tracks using the Modules for Experiments in Stellar Astrophysics (MESA\cite{Paxton2011, Paxton2018}{}) code (r10398). We adopt an exponential diffusive overshooting prescription\cite{Herwig2000c}{} with a convective-core and non-burning shell overshooting value of $f_{\rm ov} = 0.02$ expressed in pressure scale heights, initial hydrogen and metal mass fractions of $X = 0.71$ and $Z = 0.02$, respectively, OP opacity tables\cite{Seaton2005}{} and the chemical mixture of Nieva \& Przybilla \cite{Nieva2012, Przybilla2013b}{}, from the start of hydrogen core burning (that it the zero-age main sequence) to the end of core-helium burning. Adiabatic eigenfrequencies of these models were calculated for dipole zonal modes with radial orders for g modes ($n \in [-50, -1]$) and p modes ($n \in [1, 10]$) using the pulsation code GYRE\cite{Townsend2013b, Townsend2018a}{}.


\subsection*{Data availability}
The data that support the plots within this paper and other findings of this study are available from the corresponding author upon reasonable request.

\subsection*{Code availability}
The K2 systematics correction code \texttt{k2sc} is freely available and documented at \url{https://github.com/OxES/k2sc}. The iterative pre-whitening code is freely available and documented at \url{https://github.com/IvS-KULeuven/IvSPythonRepository}. The \texttt{Python} Markov chain Monte Carlo code \texttt{emcee} is freely available and documented at \url{http://dfm.io/emcee/current/}. The stellar evolution code, \texttt{MESA}, is freely available and documented at \url{http://mesa.sourceforge.net/}, and the stellar pulsation code, \texttt{GYRE}, is freely available and documented at \url{https://bitbucket.org/rhdtownsend/gyre/wiki/Home}.



	
\end{document}